# Toward a Science of Autonomy for Physical Systems: Aerial Earth Science

M. Ani Hsieh
mhsieh1@drexel.edu
Drexel University

Srikanth Saripalli
srikanth.saripalli@asu.edu
Arizona State University

Gaurav Sukhatme
gaurav@usc.edu
University of Southern California

Vijay Kumar
kumar@seas.upenn.edu
University of Pennsylvania

Unmanned Aerial Vehicles (UAVs) equipped with LiDAR, electro-optical and infrared cameras, SAR and atmospheric sensors have transformed the way we acquire high spatio-temporal resolution data. For example, UAVs equipped with these sensors have been able to obtain topography at resolutions of less than one meter, revolutionizing earth sciences. Surface processes act at spatial scales on the order of a meter to produce intricate landforms and UAVs equipped with these sensors are able to measure the three dimensional spatio-temporal geometry of the earths surface and overlying anthropogenic features and vegetation at resolutions appropriate to document these processes. In addition, surface changes due to erosion, transport and sedimentation, as well as earthquakes, landslides, volcanoes can be quantified with this data.

Other examples include using UAVs to conduct damage assessment in the Philippines after Typhoon Haiyan, take census of destroyed buildings and identify flooded areas in Haiti after Hurricane Sandy, and provide aerial support for UN peacekeeping efforts in the Democratic Republic of Congo [1]. UAVs come in a wide variety of sizes which determines their flight times and payload capacities. While UAVs are often piloted remotely by one or many human pilots, they also vary in their capacity for autonomous flight [2]. Coupled with satellite imaging, unmanned and autonomous aerial vehicles are poised to transform how we monitor the millions of physical, chemical, and biological processes on planet Earth. Processes in which the amalgamation drives climate-weather patterns and geophysical phenomena across the globe and affects food-production, contaminant and disease dispersion, natural disasters, and various other activities of significant economic and ecological impact.

---

## Emerging Fields and Application Areas

Hobby-grade unmanned aerial vehicles have become wide-spread, primarily for recreation and aerial photography. For example, over 60,000 Parrot AR.Drones have been sold in the third quarter of 2012 alone [3]. More recently, companies such as Amazon, Google have launched high-profile UAV projects. The global unmanned aerial vehicle market is forecast to reach USD 12bn by 2025 [4] with the U.S. military UAV market expected to generate more than USD 80bn in revenue in the next three years alone [5]. As UAV technologies mature and FAA finalizes the opening of the airspace to unmanned and autonomous aerial vehicles (UAVs and AAVs), the interest in deploying these systems for various scientific, commercial, and civilian applications is increasing exponentially. Applications include:

• Providing inexpensive wireless broadband (Google Loom)

• Package delivery (Amazon)

• Fighting forest fires

• Surveillance of key civil infrastructures, e.g., highways and bridges

• Structural health monitoring of oil/gas pipelines, buildings, wind turbines, bridges, and high- ways

• Delivery of medical supplies to inaccessible war torn regions, e.g., Syria

As computing, sensing, and communications technologies improve and mature and autonomy improves, AAVs will become one of the most versatile sensing assets in our global environmental monitoring network. Different satellites and stationary sensors, these aerial vehicles provide complementary sensing given their easily reconfigurable sensor payload. In addition, their mobility and autonomy allow them to obtain measurements in regions not covered by satellites and other stationary sensors over long periods of time. When coupled with satellites and stationary sensors, it is possible to achieve large scale high resolution spatio-temporal resolution sampling of the atmosphere and the Earths land and ocean surfaces.

Such capabilities will enables better monitoring and assessment of the health of our forests, prairies, wetlands, rivers, and oceans; endangered and diseased animal populations, their migratory patterns and their impact on the spread of diseases; the onset of earthquakes, volcanic eruptions, land- and mud-slides, and avalanches; and climate-weather patterns across the globe. The information gleaned can be used to refine AAVs monitoring trajectories such that they continuously adapt to changing environmental conditions. The ability to better model, track, and forecast the various geophysical processes that impact our day-to-day lives will enable us to minimize the human and economic impact of natural disasters, provide better stewardship of Earths limited natural resources, and ensure the safety of our airspace.

## AAVs for Natural Resource Management

Regular land cover assessment allows detection of changes that are important for land management agencies such as the Bureau of Land Management (BLM) and the Natural Resources Conservation Service (NRCS) as well as similar state agencies. Additional changes in infrastructure such as dams and irrigation canals can also be detected, especially when Digital Elevation Models (DEMs) from UAVs are employed with decimeter resolution. Agencies in charge of infrastructure seldom get to use temporal images of these structures so they can be evaluated if safety concerns might be an issue.

Watershed and rangeland studies require DEMs along with classification of vegetation type, bare soil and the relative positioning of these two types of land cover. Hydrological modeling requires the identification of different types of vegetation and their relationship to stream channels and contributing areas. Rangeland health assessment requires the measurement of bare soil gaps between vegetation patches. All these require resolutions of the order of 10-30cm; which can easily be obtained using autonomous aerial vehicles.

Forest applications of unmanned and autonomous aircrafts have become of widespread interest and include forest fire monitoring and recovery, and forest condition assessment. The National Park Service Complimentary to forest health monitoring and assessment, vehicles can also be deployed to monitor and track populations of endangered or invasive animal or plant or insect species. The population changes in such flora and fauna populations can be monitored and tracked using a aerial data provided by aerial vehicles coupled with stationary cameras and sensors strategically placed in the natural habitat. Better estimation of forest flora and fauna populations enables the development of data-driven conservation and mitigation policies that are better targeted for the specific species.

## AAVs for Agriculture and Farming

Increased globalization and international competition, coupled with the impact of climate-change, is resulting in new social, economic, and environmental challenges for the agricultural and farming industries. Many of these challenges can be ameliorated by improved sensing and monitoring of farm lands, crops, and livestocks. In fact, UAVs have been deployed with hyperspatial digital photography in Hawaii to predict coffee bean ripeness and in California to map crop vigor in vineyards. Other examples include the use of UAVs for targeted spraying of crops and the use of visible, thermal, and multispectral sensors on an unmanned helicopters for precision viticulture applications.

Precision and high yield agriculture can leverage AAV technology to provide long term and persistent monitor both the health and the yield of crops. Ongoing research in this area includes identifying and monitoring crop stress levels to better manage water usage throughout the farm. This requires identifying variations in plant color in the fields and correlating the plant coloration to soil conditions. Coupled with atmospheric forecasting capabilities, AAVs can further mitigate the spread of air-borne high-risk plant pathogens. Another area of interest includes estimating crop yield to better anticipate labor needs at harvest time.

Additionally, AAVs can be used to monitor the well-being of livestocks and ensure the proper allocation of livestock to pastures to maximize the health of animals and lands. Aerial, possibly coupled with ground, autonomous vehicles can assist in shepherding livestock to ensure they do not wander into dangerous regions and stay within the boundaries of the farm. Employing aerial and other robotics technology for agriculture and farming applications will significantly improve our ability to maintain both the economic as well as ecological standards in the production chain of our food supply [6].

## AAVs for Oceanographic and Atmospheric Sciences

The Deep Water Horizon and recent California Oil Spill and the crash of the Malaysian Airlines plane in the Indian Ocean has seen the use of UAVs to track the dispersion of oil and wreckage. Recently, UAVs have assisted in the collection of pollen samples to validate atmospheric models used to predict pollen propagation across large geographical regions. The ability to forecast and predict the dispersion of pollen helps us better understand the impact of invasive plant species on native flora and fauna as well as concerns raised on cross pollination between genetically and non-genetically modified crops.

The understanding of the air-sea interface is crucial for improving our weather and ocean forecasting capabilities. AAVs can be deployed to work in concert with surface drifter sensors and autonomous vehicles to better understand the energy exchange between the atmosphere and the ocean, which is crucial for improving predictions for the strength and expected path of a hurricane. Similarly, AAVs can be integrated with sensor networks on the ground to better measure Eddy covariance, soil moisture content, and ground heat flux which directly impacts the health of farmlands, wetlands, urban areas. Furthermore by coupling surface and sub-surface measurements we can better understand plant phenology and stream direction and the impact of climate-change flora and fauna health and diversity.

AAVs have the potential to significantly increase both the amount of and the spatial and temporal resolution of the data that can be gathered from our oceans and atmosphere. The ability to better predict climate weather patterns across the globe can transform how we mitigate the impact of weather related natural disasters. The ability to better predict the duration and severity of droughts, the strengths and trajectories of tropical storms, and track changes in water levels in rivers and lakes can significantly improve our ability to manage our water and flood control resources, food production capabilities, and response to natural disasters.

## AAVs for Safety, Security, Rescue, and Humanitarian Needs

Recent years has seen the increased use of UAVs in search and rescue and humanitarian applications. UAVs have been deployed in the Philippines after Typhoon Haiyan for damage assessment, in Haiti after Hurricane Sandy to take census of destroyed infrastructure and identify flooded regions, and in the Democratic Republic of Congo for UN peacekeeping efforts [1]. The difference between the pre-Sandy satellite imagery data and the post-Sandy images provided by the UAV's onboard camera provided a good census of damages to buildings and displacement of people. Similarly, AAVs can assist in locating stranded survivors in the aftermath of a natural disaster, similar to those stranded by Hurricane Katrina, or survivors trapped by landslides, earthquakes, or avalanches using infrared and/or penetrating imaging sensors. Different from UAVs, AAVs have ability provide persistent and continuous monitoring of disaster stricken regions and result in faster recovery time of stranded or trapped survivors.

In Syria, UAVs were crucial in the delivery of much needed medicine in regions inaccessible to health workers. In the Republic of Congo, UAVs provided aerial support and assisted UN Peace- keeping forces in preventing identifying, locating, and preventing rebel activities in protected regions. Similarly, AAVs can provide aerial support and assistance to law enforcement and homeland security agencies to monitor and ensure the safety of our highways, railways, shipping lanes, and airspace. Autonomous aerial systems can also provide assistance in the surveillance of borders, power plants, nuclear plants, and pipelines. And AAVs can assist in the safety monitoring of urban environments such as parks, buildings, tourist sites and in the management of crowds and maintenance of civil law.

Crucially important to the development of AAV and robotics and automation technology for safety, security, rescue, and humanitarian needs is the development of appropriate legal and regulatory policies. In particular, privacy concerns in regards to the amount of data collected and processed by law enforcement and third party entities must be addressed to ensure the protection of civil liberties. The UN's Office for the Coordination of Humanitarian Affairs recommends prioritizing transparency and community engagement in the development of best practices when addressing privacy concerns, codes of conduct, and data security guidelines for AAVs in humanitarian efforts [1]. Regulations must also tackle issues of liability and safety, especially when vehicles must operate in close proximity to civilians, bridges, roadways, etc. And humanitarian and non-governmental organizes operating in conflict zones must address ethical and operational considerations that may threaten their neutrality [1].

## Challenges and Opportunities

UAVs have the potential to significantly increase both the amount of data and the spatial and temporal resolution that can be gathered to enable better decision making in our stewardship of our limited natural resources and the safety and security of our airspace. To achieve this, the following scientific and technological challenges and improvements must be addressed:

- Distributed sensing and sensor fusion with disparate air and ground sensors

    While current UAVs are capable of autonomous flight, autonomy is limited and the vehicles are inherently limited by their sensors and their flight dynamics. A single UAV can sense a limited region; a distributed swarm of autonomous aerial vehicles combining various disparate sensors (air and ground) can "map" an entire region effectively and allow researchers to make informed decisions. Algorithms and protocols that are able to combine various low and high bandwidth sensors flying on multiple AAVs with various delays need to be developed.

- Safety (Collision avoidance, collision with critical civil infrastructure, commercial airlines, etc.)

    There is a strong need to integrate UAVs and AAVs into the national airspace due to there expanded applications in civil domains. "Sense and Avoid" capability is one of the technological challenges for such integration. In "Sense and Avoid", an vehicles must be able to autonomously detect and avoid collisions with aircraft and other obstacles in the air. Research into integrating automatic dependent surveillance-broadcast (ADS-B), ground based radar, onboard sensors, and autonomous path planning and collision avoidance needs to be performed.

- Energy efficiency of vehicles (control, chassis design, battery technology, kinematics and dynamics)

    Currently most UAVs can be classified into fixed wing, vertical take-off and landing (VTOL) or a combination of both. With NASA thinking of exploring Europa and Mars with unmanned and autonomous aerial vehicles, these vehicles are already being used for volcanic exploration. There needs to be further research on various other kinds of AAVs with varying flight and mobility modalities. Specifically questions, such as the trade-offs between efficiency, robustness against disturbances, and agility need to be answered. Furthermore the theoretical limits, and practically achievable vehicle(s) that can be designed and built by optimizing control strategies and mechanical design need to be understood.

    These issues are of particular importance since the environments they operate in more significantly impact smaller vehicles. On the other hand, the environment may be leveraged to improve vehicle control and autonomy particularly for aerial vehicles. Consider the examples of a micro-UAV or micro-AAV flying outdoors on a gusty day. The small vehicle sizes result in more tightly coupled vehicle and environmental dynamics. While this coupling of the vehicle and environment dynamics makes

planning and control challenging, the environmental forces can be exploited to extend the power budgets of small, resource-constrained vehicles, effectively prolonging their lifespan.

Biology is leading to a new class of flapping design and insect-scale aerial vehicles. Bio-inspiration and bio-mimicry has the potential to provide new energy-efficient vehicle designs and control paradigms. Biology have also shown how it is possible to leverage environmental forces to generate energy efficient control strategies and vehicle trajectories.

- Large scale data assimilation to climate, ocean, geological models

    Unmanned as well as manned aerial vehicles with EO/IR, LIDAR and SAR sensors are producing enormous amounts of data (GBytes of data/flight). Dazzling visual displays of wind and ocean current data assimilated from satellite, stationary ground, and aerial sensors have been achieved (NASAs Perpetual Ocean and http://earth.nullschool.net/). However, this data needs to be processed and higher level products such as Digital Elevation models, GIS products, and atmospheric currents need to be produced rapidly for researchers to act on this.

    Real time mapping, modeling and data processing algorithms need to be developed. Rapid data gathering and data processing must also be complemented by rapid dissemination, integration with contextual data, and the preparation of well-documented higher order data products to make the observations most useful. These algorithmic advances must be done in concert with the development of appropriate geophysical models of the environment. A major challenge in this endeavor is the immense complexity of the atmospheric and the ocean dynamics, which involves the interplay of rotation, stratification, complex topography and variable thermal and atmospheric/oceanic forcing, not to mention thousands of biological, chemical, and physical inputs.

    Theoretical and experimental efforts to model atmospheric and ocean flows have made progress with the help of simpler, so-called "reduced" models, but these models are often scale invariant.

On the hand, atmospheric and ocean current forecasts provided by National Oceanic and Atmospheric Administration (NOAA) and other agencies is assimilated from satellite and field observations coupled with predictions from numerical models. So accessibility to and the overall quality of this data is highly dependent on how well a given region of interest is instrumented, if at all. As such, algorithmic advances in data processing, assimilation, and predictions must go hand in hand with the development of the appropriate environmental models.

- Supervisory Control - one to many control of AAVs (single operator multiple AAVs)

    Existing UAVs require different human operators to monitor the various live sensor feeds, taking, and control. However, as many more vehicles are deployed, we need to pursue a command and control paradigm where single human operators can easily monitor, command, control, and re-task multiple assets. This is especially important as autonomy of single vehicles become more robust and simultaneous deployment of multiple vehicles can be streamlined. New developments in human-robot and human-computing interfaces and in control architectures will be needed. These efforts must be complemented with developments in new abilities for extracting, synthesizing, and presenting data fused from a collection of asynchronous sensors.

- Privacy and data security

    The increased scale of ubiquitous data collection by autonomous vehicles in all aspects of our everyday lives will enable better decision making to ensure long-term sustainability of our economy, environment, and safety and security. However, these capabilities also pose significant technological, political, and societal challenges in privacy and data security. Investment in the scientific research and development of suitable technology in data privacy and security must go hand in hand with regulatory and policy efforts.

## Summary and Conclusions

Unmanned Aerial Vehicles equipped with ranging, electro-optical and infrared cameras, SAR and atmospheric sensors will transform the way we identify, monitor, and interact with the millions of physical, chemical, and biological processes on planet Earth that impacts our day-to-day lives. UAV technologies can significantly improve our land and natural resource management, search and rescue operations, manage conflict in politically unstable regions, and provide humanitarian assistance in the event of man-made or natural disasters. The ability to increase the data gathered through persistent monitoring of our environments will enable better decision making in all these application spaces. These are all activities with significant economic and ecological impact that are crucial for the long-term sustainability of life as we know it.

While the increased scale of ubiquitous data collection by autonomous vehicles in all aspects of our everyday lives will pose significant challenges in privacy and data security. The time to discuss the impact of these technologies on is now. The Stimson Task Force on UAV Policy has iterated the importance of a realistic view of UAVs, and consequently AAVs, and recognizing their capabilities and the new uses and policies they enable [7]. As such, we must address the technological challenges to safeguard the privacy, safety, and security of our citizens and our natural, institutional, and economic resources through the development of appropriate cost-benefit analysis, robust oversight and accountability mechanisms, and best practices for the myriad of applications. In conjunction, we should foster continued assessment of the relevant technologies and develop an interagency research and development strategy for autonomous aerial systems and technologies in a manner consistent with our values to advance our civil, political, and social interests [7].

*For citation use*: Hseih M. A., Saripalli S., Sukhatme G., & Kumar V. (2015). *Toward a Science of Autonomy for Physical Systems: Aerial Earth Science*: A white paper prepared for the Computing Community Consortium committee of the Computing Research Association. http://cra.org/ccc/resources/ccc-led-whitepapers/

*This material is based upon work supported by the National Science Foundation under Grant No. (1136993). Any opinions, findings, and conclusions or recommendations expressed in this material are those of the author(s) and do not necessarily reflect the views of the National Science Foundation.*